\documentclass[12pt]{iopart}

\usepackage{graphicx}

\begin{document}

\title{Sampling electronic structure QUBOs with Ocean and Mukai solvers}

\author{Alexander Teplukhin$^1$, Brian K. Kendrick$^1$, Susan M. Mniszewski$^2$, Sergei Tretiak$^1$ and Pavel A. Dub$^3$}

\address{$^1$ Theoretical Division (T-1, MS B221), Los Alamos National Laboratory, Los Alamos, New Mexico 87545, USA}
\address{$^2$ Computer, Computational and Statistical Sciences Division (CCS, MS B214), Los Alamos National Laboratory, Los Alamos, New Mexico, USA}
\address{$^3$ Chemistry Division (C-IIAC, MS K558), Los Alamos National Laboratory, Los Alamos, New Mexico 87545, USA}

\ead{bkendric@lanl.gov}
\ead{pdub@lanl.gov}

\vspace{10pt}
\begin{indented}
\item[] \today
\end{indented}

\begin{abstract}
  The most advanced D-Wave Advantage quantum annealer has 5000+ qubits, however, every qubit is connected to a small number of neighbors. As such, implementation of a fully-connected graph results in an order of magnitude reduction in qubit count. To compensate for the reduced number of qubits, one has to rely on special heuristic software such as \textit{qbsolv}, the purpose of which is to decompose a large problem into smaller pieces that fit onto a quantum annealer. In this work, we compare the performance of two implementations of such software: the original open-source qbsolv which is a part of the D-Wave Ocean tools and a new Mukai QUBO solver from Quantum Computing Inc. (QCI). The comparison is done for solving the electronic structure problem and is implemented in a classical mode (Tabu search techniques). The Quantum Annealer Eigensolver is used to map the electronic structure eigenvalue-eigenvector equation to a type of problem solvable on modern quantum annealers. We find that the Mukai QUBO solver outperforms the Ocean qbsolv for all calculations done in the present work, both the ground and excited state calculations. This work stimulates the development of software to assist in the utilization of modern quantum annealers.
\end{abstract}

\vspace{2pc}
\noindent{\it Keywords}: quantum annealing, Ising optimization, qbsolv, electronic structure, Quantum Annealer Eigensolver (QAE)

\section{Introduction}

Adiabatic quantum computation (AQC) is a form of quantum computation, where an initial ``easy to prepare'' Hamiltonian is intentionally slowly (adiabatically) evolving to the final target Hamiltonian, the ground state of which one would like to obtain. The AQC and more popular gate-based quantum computation are two seemingly different approaches, which however were shown to be formally equivalent \cite{equiv}. While gate-based quantum computation is implemented on several types of Noisy Intermediate-Scale Quantum (NISQ) devices, D-Wave is currently the sole manufacturer of adiabatic quantum devices. D-wave annealers are yet to be full-fledged AQC systems, however, the current hardware is capable of performing Ising optimisation or, alternatively, quadratic unconstrained binary optimization (QUBO). The QUBO is the problem of minimizing a quadratic polynomial over binary variables. Namely, for a given matrix $Q$, the sought out solution is a binary string $x$ which minimizes
\begin{eqnarray}
E(x)=\sum_{i=1}^N \sum_{j=1}^i x_i Q_{ij} x_j. \label{eq:qubo}
\end{eqnarray}

Similar to their NISQ counterparts \cite{nisq}, modern annealers still possess a number of imperfections \cite{perspectives}. Consequently, the development of the software which bears the burden of handling those, is of the utmost importance. Specifically, the D-Wave 2000Q and Advantage architectures operate with 2048 and 5640 qubits, respectively. This may seem a lot, however, the connectivity is an issue. A single qubit is connected correspondingly to 6 and 15 neighboring qubits on these two devices. Stated differently, not every possible pair of $x_i$ and $x_j$ is present in \eref{eq:qubo} when minimization is performed on a quantum annealer. A commonly-used technique called \textit{chaining} solves this problem, where a chain of qubits from the source qubit $A$ to the target qubit $B$ is formed so that $A$ and $B$ can ``talk''. Effectively, each logical variable maps onto a chain of physical qubits. The downside of this technique is that the size of resulting fully-connected graph (or network) of QUBO variables is very small, being 64 and 124 variables on the D-Wave 2000Q and Advantage, respectively.

A way to overcome this small-size limitation of fully-connected QUBOs, is to decompose a given large QUBO problem into pieces, called subQUBO. The latter are then optimized separately on a quantum annealer aiming to construct a solution to the large QUBO problem. Multiple iterations of this procedure with a wisely chosen decomposition scheme are expected to ultimately converge to a good approximate (if not exact) solution to the initial QUBO problem. This algorithm underpins what qbsolv \cite{ocean-qbsolv} software does. Namely, the code starts with a random binary string initial guess for a solution, sorts out all qubits in terms of ``importance'' (based on bit-flip cost), splits the sorted array into pieces, sends them to a quantum annealer, and collects back the optimal binary sub-strings. The latter allow for the construction of a new total solution approximation and then start a subsequent iteration. Qbsolv also refines the global solution between iterations using Tabu search \cite{tabu} to improve the solution quality and, by doing so, does many internal restarts to exhaust improvement options. Qbsolv is an open-source code being a part of the D-Wave Ocean tools \cite{ocean}. Recently, qbsolv has been reimplemented by Quantum Computing Inc. The now-proprietary QUBO solver is focused on superior classical performance and is provided via the cloud as a part of the Mukai environment \cite{mukai}. The new sampler is parallel, exploits advanced Tabu search techniques and was recently tested on a number of problems \cite{qci-qbsolv1, qci-qbsolv2}.The  Mukai platform is quantum-ready and targets solving constrained-optimization problems.

In this work, we compare the two QUBO solver implementations, the open-source qbsolv (referred to as Ocean qbsolv) and the new proprietary version (referred to as Mukai QUBO solver). We focus on the electronic structure problem, where the Quantum Annealer Eigensolver (QAE) is used to map the electronic structure eigenvalue equation to QUBO problems, which are in turn solved using either the qbsolv or Mukai QUBO solver. Previously, the QAE was shown to work for a variety of theoretical chemistry tasks including vibrational \cite{qae-vib}, scattering \cite{qae-scat} and electronic \cite{qae-el} problems. Since a substantial body of results was obtained in the latter study \cite{qae-el}, we will use the electronic/QAE setup as a representative case study to compare the two versions of qbsolv. The comparison is completely classical, because the Mukai QUBO solver is purely classical at the present moment (i.e., all subQUBOs are solved on the CPU). Quantum Processing Units (QPUs) will be supported in an imminent Mukai release \cite{qci-qbsolv2}. In the next section, we will briefly cover the electronic structure QUBO framework. The complete procedure description can be found in the Methods section of \cite{qae-el}.

\section{Electronic structure QUBOs}

A typical goal of electronic structure calculations is to obtain the ground and sometimes a few excited states of a given molecule at a fixed nuclear configuration (i.e., geometry), solving for their energies and wavefunctions. Mathematically, the latter are eigenvalues and eigenvectors, respectively, of the electronic Hamiltonian operator, which can be represented with a matrix that describes a molecule to a desired degree of discretization (numerical grid or basis set) \cite{szabo}. Thus, one typically constructs such a matrix and then diagonalizes it or, alternatively, uses matrix-free methods to obtain a desired number of eigenpairs. The representative approaches are full configuration interaction (FCI) and complete active space self-consistent field (CASSCF), to name a few. The FCI ultimately provides an exact solution to the electronic problem (subject to basis set limitation), whereas CASSCF is exact only within the chosen active orbital subspace being a practical approximation to the FCI method. In order to compute eigenpairs on a D-Wave quantum annelear, we construct the Hamiltonian matrices explicitly and use the QAE to map the eigenvalue problem to the QUBO formulation \eref{eq:qubo}.

To construct FCI and CASSCF matrices in a given basis set \cite{szabo}, we use an in-house modified Psi4 code \cite{psi4} with the Cartesian coordinates of atoms of selected molecules optimized at the Hartree-Fock level as input data. Once the matrix is constructed and printed to a file, the QAE is called to process the matrix file. The QAE is based on the min-max variational theorem, where the Rayleigh-Ritz quotient (RRQ)
\begin{eqnarray}
R_A=(v,Av)/(v,v) \label{eq:rrq}
\end{eqnarray}
is minimized over vectors $v$ for a given matrix $A$. The minimum of the RRQ is the smallest eigenvalue, whereas the optimal vector $v_{opt}$ is the corresponding eigenvector. The RRQ expression does look similar to the QUBO form \eref{eq:qubo}, especially if $v$ happens to be normalized. However, a few additional steps should be in place to make them fully compatible. First, the elements of $v$ need to be encoded in terms of binary variables $x_i$. The QAE implements a power-of-two encoding for that. Second, ratios, such as \eref{eq:rrq}, are not allowed in the form \eref{eq:qubo}, so the RRQ ratio is replaced with a weighted sum of the numerator and the denominator giving the final objective function
\begin{eqnarray}
F(v)=(v,Av)+\lambda\cdot(v,v) . \label{eq:f}
\end{eqnarray}
The unknown Lagrangian multiplier $\lambda$ is found iteratively. This is the reason why multiple QUBO problems have to be generated and solved to obtain an eigenpair, not just a single QUBO problem. The iterative procedure for $\lambda$ avoids or discourages the trivial solution $v=\mathbf{0}$ (null vector). Moreover, equation \eref{eq:f} is another way of presenting the eigenvalue problem $Av=Ev$ (i.e., here taking a dot product with $v$ and moving $E(v,v)$ to the left of the equal sign results in the form \eref{eq:f}). As such, the optimal $\lambda_{opt}$ ends up being near (or equal) to the negative of the true eigenvalue $E$. Once $\lambda_{opt}$ and $v_{opt}$ are found, the eigenvalue is evaluated as $(v_{opt},Av_{opt})$. Other eigenpairs can be found in a similar fashion, with the input matrix appropriately modified to shift previously computed eigenvalues to lie higher in the eigenspectrum (Brauer's theorem \cite{brauer}) and applying the described procedure again without any changes, see \cite{qae-el} for more details.

In the present case study, all of the generated QUBOs are one to two orders of magnitude larger than the fully-connected QUBOs supported by modern D-Wave quantum annealers. This is where the qbsolv and Mukai QUBO solvers come to the rescue through their ability to handle large QUBO problems.

\section{Results}

In what follows, we compare the accuracy of the two QUBO solvers (classical mode with all subQUBOs solved on the CPU). Similar to the calculations done in the previous work \cite{qae-el}, we start with the H$_2$ ground state energy evaluation and study its convergence with basis set. Specifically, we use 14 different basis sets with increasing size. Consequently this leads to 14 FCI matrices sized from 2x2 (STO-3G) to 1256x1256 (aug-cc-PVQZ). When applying the same level of discretization of eigenvector elements $K=10$, the resulting QUBOs have 20 to 12560 variables, respectively.

\begin{figure}
\centering
\includegraphics[scale=1.0]{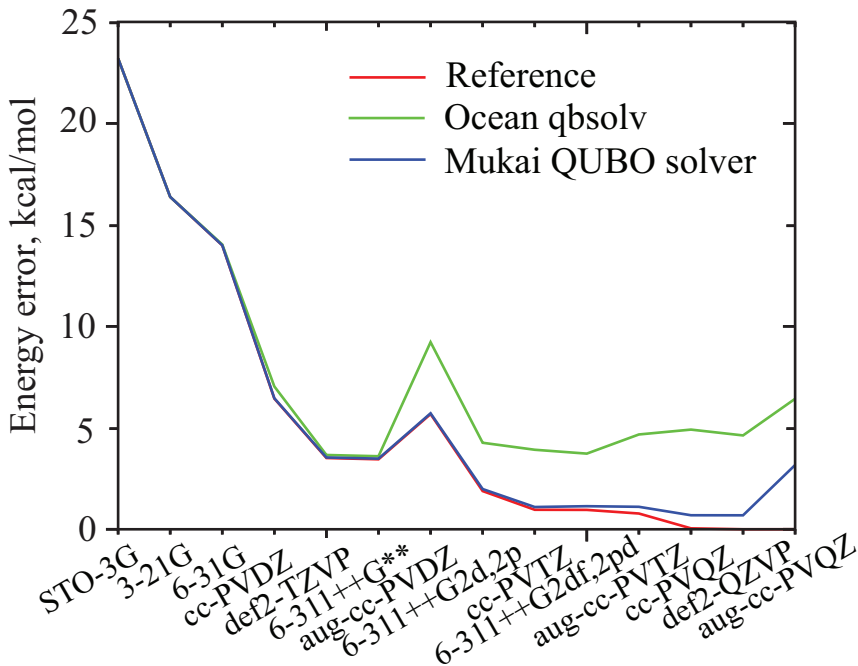}
\caption{\label{fig01} Convergence of the H$_2$ ground state energy with basis set expansion. The calculations are ordered with increasing Hamiltonian matrix size. The matrices were diagonalized directly (red) and using the QAE + Ocean qbsolv (green) or QAE + Mukai QUBO solver (blue) approaches. The error is given relative to the energy of def2-QZVP reference diagonalization, which has the lowest value among all calculations. The Mukai QUBO solver outperforms the Ocean qbsolv throughout all basis sets by a significant margin.}
\end{figure}

\Fref{fig01} illustrates the calculated energy convergence with basis set. The blue and green curves show the Mukai QUBO solver and Ocean qbsolv results, respectively. The reference calculation is an exact matrix diagonalization using SciPy/LAPACK \cite{lapack} (the red and green curves are the same as in Figure 1 of \cite{qae-el}). We observe that the QAE + Mukai QUBO solver combination substantially outperforms the QAE + Ocean qbsolv in terms of accuracy. The classical noise floor for the Mukai QUBO solver (i.e., the part of the curve where the error stopped decreasing) appears much later and is significantly lower than that for the Ocean qbsolv. Almost all errors in the blue curve are within chemical accuracy of 1 kcal/mol \cite{kcal}. The only exception is the last point, where the largest matrix is involved. Still, even in this case with 12560 QUBO variables the Mukai QUBO solver error 3 kcal/mol is twice smaller compared to Ocean's 6 kcal/mol error.

\Tref{table1} lists the ground state energy errors for the different molecular species considered in the Table 1 of our previous work \cite{qae-el}. Again, the Mukai QUBO solver shows much smaller errors compared to the Ocean qbsolv values, being 10 to 100 times smaller on average. Importantly, all Mukai errors are less than 1 kcal/mol, which means that the Mukai QUBO solver gives results to within chemical accuracy. To compare the quality of the computed ground state wave functions, we add two more columns to the table with the norms of eigenvector residuals calculated as $R_{solver}=\sum_i|v_i^{solver}-v_i^{ref}|$, where $solver$ is either the Ocean qbsolv or Mukai QUBO solver and the summation runs over the eigenvector elements. As with the energies, QAE + Mukai gives more accurate wave functions, i.e. smaller residual norms, than QAE + Ocean. The $R_{mukai}$ values tend to be 2 to 10 times smaller than the $R_{ocean}$ values.

\begin{table}
\caption{\label{table1} Electronic ground state energy errors in kcal/mol.}
\fontsize{10}{12}\selectfont
\begin{tabular}{@{}ccccccccc}
\br
Molecule & Method & Basis & Mat. size & QUBO size & $\Delta E_{ocean}^{\rm a}$ & $\Delta E_{mukai}^{\rm b}$ & $R_{ocean}^{\rm c}$ & $R_{mukai}^{\rm d}$ \\
\mr
H$_2$       & FCI           & STO-3G     & 2x2       & 20    &  0.000  & 0.000 & 0.000 & 0.000 \\
HF          & FCI           & STO-3G     & 18x18     & 180   &  0.152  & 0.017 & 0.017 & 0.007 \\
H$_2$O      & FCI           & STO-3G     & 133x133   & 1330  &  5.484  & 0.084 & 0.268 & 0.042 \\
H$_2$O      & CAS(8e,7o)SCF & cc-PVDZ    & 321x321   & 3210  &  8.912  & 0.123 & 0.626 & 0.089 \\
CH$_2^{2+}$ & FCI           & STO-3G     & 169x169   & 1690  &  5.265  & 0.071 & 0.344 & 0.036 \\
BeH$_2$     & FCI           & STO-3G     & 169x169   & 1690  &  2.079  & 0.043 & 0.258 & 0.040 \\
H$_3^+$     & FCI           & cc-PVTZ    & 532x532   & 5320  &  6.934  & 0.460 & 0.452 & 0.161 \\
BH$_3$      & CAS(6e,6o)SCF & 6-311++G** & 208x208   & 2080  &  6.673  & 0.086 & 0.550 & 0.081 \\
BH$_3$      & FCI           & STO-3G     & 1250x1250 & 12500 & 10.512  & 0.510 & 0.685 & 0.239 \\
\br
\end{tabular}
\fontsize{8}{9.6}\selectfont
$^{\rm a}$ Energy difference between the QAE + Ocean qbsolv ($E_{ocean}$) and the reference diagonalization ($E_{ref}$) \cite{qae-el}. \\
$^{\rm b}$ Energy difference between the QAE + Mukai QUBO solver ($E_{mukai}$) and the reference diagonalization ($E_{ref}$). \\
$^{\rm c}$ Norm of eigenvector residual for the QAE + Ocean qbsolv. \\
$^{\rm d}$ Norm of eigenvector residual for the QAE + Mukai QUBO solver.
\end{table}

We also test both QUBO solvers on excited states, where several eigenpairs are computed sequentially as described in the previous section. \Tref{table2} shows four transition energies for the FCI/STO-3G water molecule (133x133 matrix), which are the spacings between the smallest eigenvalues. Similar to the ground state calculations, we find that the QAE + Mukai QUBO solver combination gives much smaller errors when compared to the QAE + Ocean qbsolv. The error is less than 1 kcal/mol for the first three transitions. The last transition  S$_0$ $\rightarrow$ S$_4$ seems to be challenging for both QUBO solvers, however the Mukai QUBO solver error 3.73 kcal/mol is still four times smaller than Ocean's. Interestingly, the Mukai error for the S$_0$ $\rightarrow$ S$_{3}$ transition happened to be extremely small, 10$^{-3}$ kcal/mol. This might due to a lucky cancellation of errors that come from previously computed eigenvectors. All QAE energies are not final and can change from run to run. The fluctuations are expected, because both QUBO solvers are heuristic. The residual norms $R$ for the four excited states of water are given in the last two columns of the table. As we can see, $R_{mukai}$ is much smaller than $R_{ocean}$ for all states, which means that the Mukai QUBO solver takes the lead in the quality of the computed wave functions as well.

\begin{table}
\caption{\label{table2} Electronic transition energies of the H$_2$O molecule computed using the QAE and FCI/STO-3G matrix in kcal/mol.}
\fontsize{10}{12}\selectfont
\begin{tabular}{@{}cccccccc}
\br
Transition & $T_{ref}^{\rm a}$ & $T_{ocean}^{\rm b}$ & $T_{mukai}^{\rm c}$ & $T_{ocean}-T_{ref}$ & $T_{mukai}-T_{ref}$ & $R_{ocean}^{\rm d}$ & $R_{mukai}^{\rm e}$  \\
\mr
S$_0$ $\rightarrow$ S$_1$ & 303.056 & 300.563 & 303.115 & -2.493 & 0.059 & 0.360 & 0.079 \\
S$_0$ $\rightarrow$ S$_2$ & 369.233 & 373.585 & 369.440 &  4.352 & 0.207 & 0.732 & 0.115 \\
S$_0$ $\rightarrow$ S$_3$ & 441.058 & 437.217 & 441.059 & -3.841 & 0.001 & 0.352 & 0.131 \\
S$_0$ $\rightarrow$ S$_4$ & 590.407 & 606.617 & 594.137 & 16.210 & 3.730 & 2.295 & 0.973 \\
\br
\end{tabular}

\fontsize{8}{9.6}\selectfont
$^{\rm a}$ Transition energy obtained using the reference diagonalization \cite{qae-el}. \\
$^{\rm b}$ Transition energy obtained using the QAE + Ocean qbsolv \cite{qae-el}. \\
$^{\rm c}$ Transition energy obtained using the QAE + Mukai QUBO solver. \\
$^{\rm d}$ Norm of eigenvector residual for the QAE + Ocean qbsolv. \\
$^{\rm e}$ Norm of eigenvector residual for the QAE + Mukai QUBO solver.
\end{table}

\begin{figure}
\centering
\includegraphics[scale=1.0]{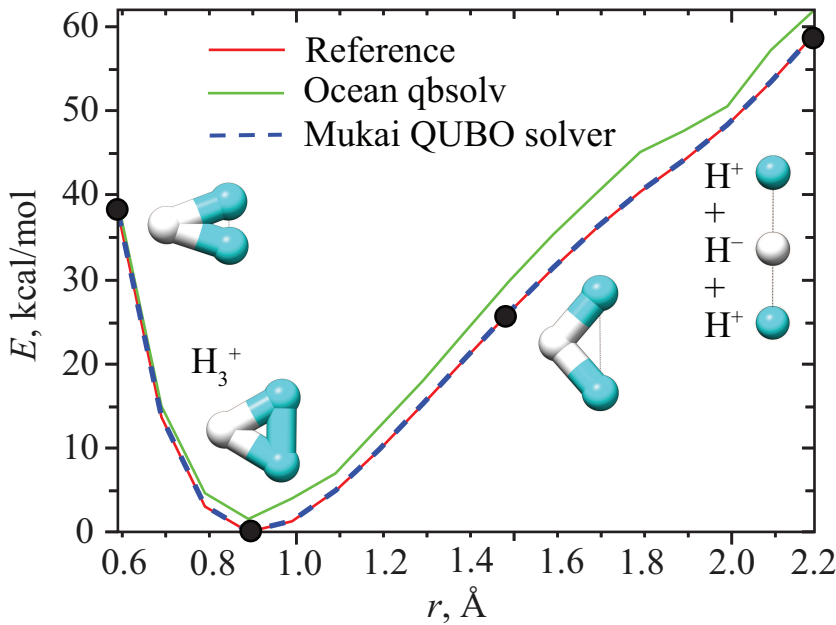}
\caption{\label{fig02} Potential energy curve of H$_3^+$ computed at FCI/cc-PVDZ level using the same methods as in Figure 1 (same colors). Molecular images (taken at black points) show how the minimum geometry evolves as a function of the distance between two terminal hydrogens. QAE + Mukai QUBO solver curve (blue) overlaps with the reference diagonalization curve and is made dashed, so that both curves are visible. }
\end{figure}

Finally, both QUBO solvers are benchmarked on the calculation of the potential energy curve of H$_3^+$, see \Fref{fig02}. In this test, 17 distinct geometries were generated at the FCI/cc-PVDZ level, each pre-optimized using the restricted Hartree-Fock method, where the distance between the two terminal hydrogens was varied from 0.59 to 2.19 \AA{} following a step size of 0.1 \AA{} (i.e., constrained optimization). Again, the Mukai QUBO solver outperforms the Ocean qbsolv. The corresponding energy curve closely follows the energy curve obtained with direct diagonalization, whereas Ocean's curve, taken from our previous study \cite{qae-el}, consistently overestimates the H$_3^+$ electronic energy. The QAE + Mukai error is less than 0.1 kcal/mol throughout the whole range of the coordinate. In contrast, the maximum error in the case of Ocean qbsolv is 4.7 kcal/mol.

\section{Conclusions and outlook}
The development of a versatile and robust software base is imperative for the rapid adaptation of emerging quantum computing hardware. The solution of quantum chemical problems is one of the immediate applications of quantum computers. In this purely classical study, we compare two QUBO solvers, which were developed to handle large QUBO problems by decomposing them into smaller pieces that fit on the D-Wave quantum annealers. The comparison is done in the context of electronic structure calculations. An in-house modified Psi4 code was used to construct molecular Hamiltonian matrices and the QAE was then used to convert the resulting eigenvalue problems to QUBO problems. The test cases include the ground state energy convergence with basis set for the hydrogen molecule, the ground state energy calculation for different molecular species up to tetraatomics and an excited state calculation for the water molecule. We find that the Mukai QUBO solver from Quantum Computing Inc. is one to two orders of magnitude more accurate than the qbsolv from the D-Wave Ocean tools. It is important to note that almost all ground and excited state energy errors based on the Mukai QUBO evaluation fall within the chemical accuracy of 1 kcal/mol.

As a final note, we also did some tests with the D-Wave Hybrid Solver Service \cite{hss}, which is an alternative to the two QUBO solvers. Unfortunately, the current version of the QAE does not work with the HSS. There is a class of QUBO problems constructed by the QAE for which we expect to get the trivial solution $v=\mathbf{0}$ (i.e, a null vector), and both versions of qbsolv do return this specific QUBO solution. In contrast, the HSS returns a non-trivial solution corresponding to a higher QUBO minimum, which in turn breaks the QAE iteration logic. Thus, a change to the QAE may be needed to make it work with HSS. This will be explored in the future. Overall, we welcome the further development of quantum annealing hardware and emergence of respective software in the field.

\ack{Research presented in this article was supported by the Laboratory Directed Research and Development (LDRD) program of Los Alamos National Laboratory (LANL) under project number 20200056DR. This work was conducted in part at the Center for Integrated Nanotechnologies, a U.S. Department of Energy, Office of Basic Energy Sciences user facility. The authors would like to thank Quantum Computing Inc. for providing a free trial of Mukai.}

\section*{References}
\bibliographystyle{unsrt} % iopart-num
\bibliography{main}

\end{document}